\documentclass[11pt]{article}

\usepackage{amsfonts,amsmath,amssymb,bm,graphicx,hyperref,a4wide,subfigure}

\begin{document}

%
%

\title{\textbf{Spin-flavor oscillations of Dirac neutrinos in matter under the influence
of a plane electromagnetic wave}}

\author{Maxim Dvornikov\thanks{maxdvo@izmiran.ru} 
\\
\small{\ Pushkov Institute of Terrestrial Magnetism, Ionosphere} \\
\small{and Radiowave Propagation (IZMIRAN),} \\
\small{108840 Troitsk, Moscow, Russia}}

\date{}

\maketitle

\begin{abstract}
We study oscillations of Dirac neutrinos in background matter and a plane electromagnetic wave. First, we find the new exact solution of the Dirac-Pauli equation for a massive neutrino with the anomalous magnetic moment electroweakly interacting with matter under the influence of
a plane electromagnetic wave with the circular polarization. We use this result to describe neutrino spin oscillations in the external fields in question. Then we consider several neutrino flavors and
study neutrino spin-flavor oscillations in this system. For this purpose we formulate the initial condition problem and solve it accounting for the considered external fields. We derive the analytical expressions for the transition probabilities of spin-flavor oscillations for different types of neutrino magnetic moments. These analytical expressions are compared with the numerical solutions of the effective Schr\"odinger equation and with the findings of other authors. In particular, we reveal that a resonance in neutrino spin-flavor oscillations in the considered external fields cannot happen contrary to the previous claims. Finally, we briefly discuss some possible astrophysical applications.
\end{abstract}


\section{Introduction}

Nowadays it is commonly believed that neutrinos possess nonzero masses
and mixing between different flavor eigenstates~\cite{Bil18}. These
properties of neutrinos result in transitions between neutrino flavors, which are called
neutrino flavor oscillations~\cite{Fan18}. Neutrino flavor oscillations
are known to happen even in vacuum, i.e., at the absence of external
fields.

As constituents of the standard model, neutrinos can interact with
other fermions, which a background matter is made of, by exchanging
virtual $W$ and $Z$ bosons. This kind of interaction, although it
is quite weak, can significantly influence the process of neutrino
flavor oscillations resulting in the resonance enhancement of the
transition probability, known as the Mikheyev-Smirnov-Wolfenstein
(MSW) effect~\cite{BleSmi13}. The MSW effect is believed to be the
most plausible solution to the solar neutrino problem~\cite{Raf96}.

Despite neutrinos are electrically neutral particles, nothing prevents
them to have nonzero magnetic moments~\cite{FujShr80,FukYan03}, which are
of a pure anomalous origin. Neutrino magnetic moments result in the particle
spin precession in an external electromagnetic field. Thus, a left
polarized neutrino, which exists in the standard model, can be transformed
to a right polarized particle, invisible to the detectors. If this
process happens within one neutrino generation, it is called neutrino
spin oscillations~\cite{FujShr80}. There is a possibility for neutrinos
to change both flavor and the polarization in an external electromagnetic
field. In this situation, these transitions are named neutrino spin-flavor
oscillations. Neutrino spin and spin-flavor oscillations were recently reviewed in Ref.~\cite{BalKay18}.

Neutrino spin-flavor oscillations were studied mainly in a constant
magnetic field, which is transverse with respect to the neutrino motion.
However, other nontrivial configurations of the electromagnetic field,
like an electromagnetic wave are of interest. This interest is inspired,
e.g., by the suggestion in Refs.~\cite{MeuKeiPia15,Fom18} to explore
the neutrino evolution in intense laser pulses. Note that the study
on the development of intense lasers in Ref.~\cite{StrMou85} was
recognized by the Nobel committee in 2018.

Neutrino spin and spin-flavor oscillations in matter and an electromagnetic
wave were previously discussed in Refs.~\cite{EgoLobStu00,DvoStu01}
Recently, we demonstrated in Ref.~\cite{Dvo18} that the results
of Ref.~\cite{EgoLobStu00} are not applicable for the description
of spin-flavor oscillations. In the present work, we continue the
study of Ref.~\cite{Dvo18}. However, besides the neutrino interaction
only with a plane electromagnetic wave~\cite{Dvo18}, now we account for the electroweak interaction of neutrinos 
with background matter. As in Ref.~\cite{Dvo18}, here we suppose
that neutrinos are Dirac particles. Despite multiple models for the
neutrino mass generation predict that neutrinos are likely to be Majorana
fermions~\cite{Kin17}, the nature of these particles is still unclear~\cite{EllFra15}. 

This paper is organized as follows. We start in Sec.~\ref{sec:INTERACT}
with the basics of neutrino electrodynamics in background matter.
Then, in Sec.~\ref{sec:SOLDIREQ}, we find the new exact solution
of the wave equation for a single neutrino mass eigenstate interacting
with matter and a plane electromagnetic wave with the circular polarization.
The obtained results are applied in Sec.~\ref{sec:SPINOSC} to describe
neutrino spin oscillations in the considered external fields. Then,
in Sec.~\ref{sec:DIAG}, we study neutrino spin-flavor oscillations
in matter and a plane electromagnetic wave, with the diagonal magnetic
moments being greater than the transition one. The opposite situation,
when the transition magnetic moment is dominant, is considered
in Sec.~\ref{sec:TRANS}. Some possible astrophysical applications
are also briefly discussed in Sec.~\ref{sec:TRANS}. Finally, in Sec.~\ref{sec:CONCL}, we summarize our results.

\section{Neutrino interaction with external fields\label{sec:INTERACT}}

In this section, we briefly recall how neutrinos can interact with
background matter and an external electromagnetic field. We consider
these interactions both in flavor and mass eigenstates bases.

Without loss of generality, we shall study the system of two massive
neutrinos $(\nu_{\alpha},\nu_{\beta})$ with a nonzero mixing. For
example, we can take that $\nu_{\alpha}\equiv\nu_{\mu,\tau}$ and
$\nu_{\beta}\equiv\nu_{e}$. These neutrinos can electroweakly interact
with background matter consisting of electrons, protons, and neutrons.
The background matter is supposed to be nonmoving and unpolarized.
Moreover, we shall take that neutrinos have nonzero magnetic moments
and can interact with the external electromagnetic field $F_{\mu\nu}=(\mathbf{E},\mathbf{B})$.

The Lagrangian for the system of these neutrinos has the form,
\begin{equation}\label{eq:Lagrnu}
  \mathcal{L} =
  \sum_{\lambda\lambda'=\alpha,\beta}\bar{\nu}_{\lambda}
  \left[
    \delta_{\lambda\lambda'}\mathrm{i}\gamma^{\mu}\partial_{\mu} -
    m_{\lambda\lambda'} -
    \frac{M_{\lambda\lambda'}}{2}F_{\mu\nu}\sigma^{\mu\nu} -
    \frac{f_{\lambda\lambda'}}{2}\gamma^{0}(1-\gamma^{5})
  \right]
  \nu_{\lambda'},
\end{equation}
where $\gamma^{\mu}=\left(\gamma^{0},\bm{\gamma}\right)$, $\gamma^{5}=\mathrm{i}\gamma^{0}\gamma^{1}\gamma^{2}\gamma^{3}$,
and $\sigma_{\mu\nu}=\tfrac{\mathrm{i}}{2}[\gamma_{\mu},\gamma_{\nu}]_{-}$
are the Dirac matrices. The mass matrix $(m_{\lambda\lambda'})$ and
the matrix of magnetic moments $(M_{\lambda\lambda'})$ are independent
in general. The matrix of the effective potentials of the neutrino
interaction with matter is diagonal in the flavor basis: $f_{\lambda\lambda'}=f_{\lambda}\delta_{\lambda\lambda'}$.
The explicit form of $f_{\lambda}$ in the electroneutral matter can
be obtained on the basis of the results of Ref.~\cite{DvoStu02}
as
\begin{equation}\label{eq:fnu}
  f_{\nu_{e}} =
  \sqrt{2}G_{\mathrm{F}}
  \left(
    n_{e}-\frac{1}{2}n_{n}
  \right),
  \quad 
  f_{\nu_{\mu},\nu_{\tau}}=-\frac{1}{\sqrt{2}}G_{\mathrm{F}}n_{n},
\end{equation}
where $G_{\mathrm{F}}=1.17\times10^{-5}\,\text{GeV}^{-2}$ is the
Fermi constant and $n_{e,n}$ are the number densities of electrons
and neutrons.

The nature of neutrinos can be revealed only if we transform the flavor
wave functions $\nu_{\lambda}$ to the mass eigenstates basis,
\begin{equation}\label{eq:nupsi}
  \nu_{\lambda}=\sum_{a=1,2}U_{\lambda a}\psi_{a},\quad(U_{\lambda a})=
  \left(
    \begin{array}{cc}
      \cos\theta & -\sin\theta\\
      \sin\theta & \cos\theta
    \end{array}
  \right),
\end{equation}
where $\theta$ is the vacuum mixing angle, which is chosen in such
a way to diagonalize the mass matrix: $(U_{\lambda a})^{\dagger}(m_{\lambda\lambda'})(U_{\lambda'b})=m_{a}\delta_{ab}$,
where $m_{a}$ are the neutrino masses. The neutrino mass eigenstates
$\psi_{a}$, $a=1,2$, are taken to be Dirac particles. In general
situation, the matrices of magnetic moments $(\mu_{ab})=(U_{\lambda a})^{\dagger}(M_{\lambda\lambda'})(U_{\lambda'b})$
and neutrino interaction with background matter $(V_{ab})=(U_{\lambda a})^{\dagger}(f_{\lambda\lambda'})(U_{\lambda'b})$
are nondiagonal in the mass eigenstates basis.

Using Eq.~(\ref{eq:nupsi}), we can rewrite the Lagrangian in Eq.~(\ref{eq:Lagrnu})
in the following way:
\begin{equation}\label{eq:Lagrpsi}
  \mathcal{L}=\sum_{ab=1,2}\bar{\psi}_{a}
  \left[
    \delta_{ab}
    \left(
      \mathrm{i}\gamma^{\mu}\partial_{\mu}-m_{a}
    \right) -
    \frac{\mu_{ab}}{2}F_{\mu\nu}\sigma^{\mu\nu} -
    \frac{V_{ab}}{2}\gamma^{0}(1-\gamma^{5})
  \right]
  \psi_{b}.
\end{equation}
One can see that the Dirac equations for different mass eigenstates,
resulting from the Lagrangian in Eq.~(\ref{eq:Lagrpsi}), are coupled
due to the presence of external fields.

\section{Solution of the Dirac-Pauli equation\label{sec:SOLDIREQ}}

In this section, we study the evolution of a single neutrino mass
eigenstate in matter under the influence of a plane electromagnetic
wave. We write down the wave equation for a massive neutrino in these
external fields and find its exact solution.

In this section, we neglect the mixing between different neutrino
types. Thus, using Eq.~(\ref{eq:Lagrpsi}) and omitting the index
$a$ there, we obtain the wave equation for a Dirac neutrino with
the nonzero mass $m$ and the magnetic moment $\mu$, interacting
with nonmoving and unpolarized background matter and with the external
electromagnetic field, in the form,
\begin{equation}\label{eq:DirEqLarg}
  \left[
    \mathrm{i}\gamma^{\mu}\partial_{\mu}-m-\frac{\mu}{2}F_{\mu\nu}\sigma^{\mu\nu}-  
    \frac{V}{2}\gamma^{0}(1-\gamma^{5})
  \right]
  \psi=0,
\end{equation}
where $\psi$ is the neutrino bispinor. The effective potential $V$
can be obtained basing on Eqs.~(\ref{eq:fnu}) and~(\ref{eq:nupsi}).
We suppose that a neutrino interacts with a plane electromagnetic
wave. Neglecting the dispersion of the wave, one gets that the electric
and magnetic fields, $\mathbf{E}$ and $\mathbf{B}$, depend on $t-z$.
If the wave propagates along the $z$-axis, one has that $\mathbf{B}=(B_{x},B_{y},0)$,
and $\mathbf{E}=(B_{y},-B_{x},0),$ where $B_{x}$ and $B_{y}$ are
linearly independent components of the magnetic field.

We rewrite Eq.~(\ref{eq:DirEqLarg}) in the Hamilton form as
\begin{equation}\label{eq:DirEqHam}
  \mathrm{i}\partial_{t}\psi=
  \left[
    -\mathrm{i}(\bm{\alpha}\nabla)+\beta m+\mu(\mathrm{i}\bm{\gamma}\mathbf{E} -
    \beta\bm{\Sigma}\mathbf{B})+\frac{V}{2}(1-\gamma^{5})
  \right]
  \psi,
\end{equation}
where $\bm{\alpha}=\gamma^{0}\bm{\gamma}$, $\beta=\gamma^{0}$, and
$\bm{\Sigma}=\gamma^{5}\gamma^{0}\bm{\gamma}$ are the Dirac matrices.
If the density of background matter is constant, we can gauge the
term $V/2$ out: $\psi\to\exp\left(-\mathrm{i}Vt/2\right)\psi$. It
is convenient to introduce the new variables $u_{0}=t-z$ and $u_{3}=t+z$.
Defining the derivatives with respect to $u_{0}$ and $u_{3}$ as
$\partial_{0}$ and $\partial_{3}$, one gets that Eq.~(\ref{eq:DirEqHam})
has the following integrals:
\begin{equation}
  -\mathrm{i}\nabla_{\perp}\psi=\mathbf{p}_{\perp}\psi,
  \quad
  2\mathrm{i}\partial_{3}\psi=\lambda\psi,
\end{equation}
where $\nabla_{\perp}=\left(\partial_{x},\partial_{y},0\right)$ and
$\mathbf{p}_{\perp}=(p_{x},p_{y},0)$. Here we assume that the background
matter is uniform.

Then we look for the solution of Eq.~(\ref{eq:DirEqHam}) in the
form,
\begin{equation}\label{eq:psi0def}
  \psi=\exp
  \left(
    -\mathrm{i}Vt/2+\mathrm{i}\mathbf{p}_{\perp}\mathbf{x}_{\perp}-
    \mathrm{i}\lambda u_{3}/2
  \right)
  \psi_{0},
\end{equation}
where $\psi_{0}=\psi_{0}(u_{0})$. The equation for $\psi_{0}$ reads
\begin{equation}
  \mathrm{i}(1-\alpha_{z})\partial_{0}\psi_{0}=
  \left[
    (\bm{\alpha}_{\perp}\mathbf{p}_{\perp})-\frac{\lambda}{2}(1+\alpha_{z})+\beta m+
    \mu(\mathrm{i}\bm{\gamma}\mathbf{E}-\beta\bm{\Sigma}\mathbf{B})-
    \frac{V}{2}\gamma^{5}
  \right]
  \psi_{0}.
\end{equation}
Note that the matrix $(1-\alpha_{z})$ is singular. Thus some of the
components of $\psi_{0}$ obey the algebraic rather than differential
equations.

It is convenient to choose the Dirac matrices in the chiral representation~\cite{ItzZub80}.
If we define $\psi_{0}^{\mathrm{T}}=(\psi_{1},\psi_{2},\psi_{3},\psi_{4})$,
then only $\psi_{2}$ and $\psi_{3}$ are independent and satisfy
the equations
\begin{align}
  \mathrm{i}\partial_{0}\psi_{2} = &
  \left[
    \frac{\lambda}{2}\frac{p_{\perp}^{2}+m^{2}}{\lambda^{2}-V^{2}/4}-\frac{V}{4}
    \left(
      1+\frac{p_{\perp}^{2}-m^{2}}{\lambda^{2}-V^{2}/4}
    \right)
  \right]
  \psi_{2}
  \notag
  \\
  & +
  \left\{
    \mu(B_{x}+\mathrm{i}B_{y})+
    \frac{V}{2}\frac{m(p_{x}+\mathrm{i}p_{y})}{\lambda^{2}-V^{2}/4}
  \right\}
  \psi_{3},
  \nonumber
  \\
  \mathrm{i}\partial_{0}\psi_{3} = &
  \left[
    \frac{\lambda}{2}\frac{p_{\perp}^{2}+m^{2}}{\lambda^{2}-V^{2}/4}+\frac{V}{4}
    \left(
      1+\frac{p_{\perp}^{2}-m^{2}}{\lambda^{2}-V^{2}/4}
    \right)
  \right]
  \psi_{3}
  \notag
  \\
  & +
  \left\{
    \mu(B_{x}-\mathrm{i}B_{y})+
    \frac{V}{2}\frac{m(p_{x}-\mathrm{i}p_{y})}{\lambda^{2}-V^{2}/4}
  \right\}
  \psi_{2}.
\end{align}
After the separation the common factor in $\psi_{2,3}$ as
\begin{equation}\label{eq:v12def}
  \psi_{2,3}=
  \exp
  \left(
    -\mathrm{i}\frac{\lambda}{2}\frac{p_{\perp}^{2}+m^{2}}{\lambda^{2}-V^{2}/4}u_{0} 
  \right)
  v_{1,2},
\end{equation}
one gets that the spinor $v=v(u_{0})=(v_{1},v_{2})^{\mathrm{T}}$
obeys the equation,
\begin{equation}\label{eq:effSchv}
  \mathrm{i}\partial_{0}v=(\bm{\sigma}^{*}\mathbf{R})v,
\end{equation}
where $\bm{\sigma}$ are the Pauli matrices and
\begin{equation}\label{eq:R}
  \mathbf{R}=\mu\mathbf{B}+\frac{V}{2}\frac{m\mathbf{p}_{\perp}}{\lambda^{2}-
  V^{2}/4}-\frac{V}{4}
  \left(
    1+\frac{p_{\perp}^{2}-m^{2}}{\lambda^{2}-V^{2}/4}
  \right)
  \mathbf{e}_{z}.
\end{equation}
Here $\mathbf{e}_{z}$ is the unit vector along the wave propagation.

Finally, using Eqs.~(\ref{eq:psi0def}) and~(\ref{eq:v12def}),
the general solution of the wave Eq.~(\ref{eq:DirEqLarg}) has the
form,
\begin{equation}\label{eq:psigensol}
  \psi=
  \exp
  \left(
    -\mathrm{i}\frac{V}{2}t+\mathrm{i}\mathbf{p}_{\perp}\mathbf{x}_{\perp}-
    \mathrm{i}\frac{\lambda}{2}u_{3} -
    \mathrm{i}\frac{\lambda}{2}\frac{p_{\perp}^{2}+m^{2}}{\lambda^{2}-V^{2}/4}u_{0}  
  \right)
  u,
\end{equation}
where
\begin{equation}\label{eq:basspinu}
  u = \frac{1}{\sqrt{N}}
  \left(
    \begin{array}{c}
      \frac{(p_{x}-\mathrm{i}p_{y})v_{1}-mv_{2}}{\lambda+V/2}\\
      v_{1}\\
      v_{2}\\
      -\frac{(p_{x}+\mathrm{i}p_{y})v_{2}+mv_{1}}{\lambda-V/2}
    \end{array}
  \right),
\end{equation}
is the basis spinor. The normalization coefficient $N$ is given by
the condition $|u|^{2}=1$.

\section{Neutrino spin oscillations\label{sec:SPINOSC}}

Now we use the general solution of the Dirac-Pauli equation, found
in Sec.~\ref{sec:SOLDIREQ}, to describe neutrino spin oscillations
in a plane wave with the circular polarization. We specify the initial
condition and find the transition probability. The obtained results
are compared with previous findings of other authors.

Before we proceed, it is convenient to replace the total wave function
$\psi$ in Eq.~(\ref{eq:psigensol}) by its projection to the subspace
of the linearly independent components $\psi_{2,3}$,
\begin{equation}\label{eq:psiproj}
  \psi\to\tilde{\psi}=\frac{1}{2}(1-\alpha_{z})\psi.
\end{equation}
The basis spinor $\tilde{u}$, which is the analogue of $u$ in Eq.~(\ref{eq:basspinu}),
takes the form, $\tilde{u}^{\mathrm{T}}=\left(0,v_{1},v_{2},0\right)$.
This spinor is automatically normalized to one if $|v_{1}|^{2}+|v_{2}|^{2}=1$.
The last condition results from the unitary dynamics of $v$ implied
by Eq.~(\ref{eq:effSchv}). The mean value of an operator $\hat{O}$
can be found as $\left\langle \hat{O}\right\rangle =\tilde{\psi}^{\dagger}\hat{O}\tilde{\psi}$. 

In order not to encumber the presentation, we discuss the situation
of a neutrino propagating along an electromagnetic wave, i.e. $\mathbf{p}_{\perp}=0$.
This case is implemented if neutrinos and an electromagnetic wave
are emitted by the same source. Then we consider a wave with the circular
polarization, i.e. $B_{x}=B_{0}\cos\left[g\omega(t-z)\right]$ and
$B_{y}=B_{0}\sin\left[g\omega(t-z)\right]$, where $B_{0}$ is the
amplitude of the wave, $\omega$ is its frequency, and $g=\pm1$ is the sign factor corresponding
to right or left polarizations.

The solution of Eq.~(\ref{eq:effSchv}) for a circularly polarized
wave has the form,
\begin{equation}\label{eq:vsol}
  v=\mathcal{U}_{z}
  \left[
    \cos(\Omega u_{0})-\mathrm{i}(\bm{\sigma}\mathbf{n})\sin(\Omega u_{0})
  \right]
  v_{0},
\end{equation}
where 
\begin{equation}\label{eq:UzOmega}
  \mathcal{U}_{z}=\exp(\mathrm{i}\sigma_{3}g\omega u_{0}/2),
  \quad
  \Omega=\sqrt{(R_{z}+g\omega/2)^{2}+(\mu B_{0})^{2}},
\end{equation}
and 
\begin{equation}\label{eq:n}
  \mathbf{n}=\frac{1}{\Omega}
  \left(
    \mu B_{0},0,R_{z}+g\omega/2
  \right),
\end{equation}
is the unit vector, $R_{z}$ is the $z$-component of the vector $\mathbf{R}$
in Eq.~(\ref{eq:R}), and $v_{0}$ is the initial spinor corresponding
to $u_{0}=0$.

We suppose that, at $u_{0}=0$, only left polarized neutrinos are
presented in the space-time region outside the wave propagation. Thus,
we impose the condition $\Sigma_{z}\tilde{u}=-\tilde{u}$ on the basis
spinor $\tilde{u}$ at $u_{0}=0$. The components of the spinor $v_{0}$
are $v_{1}(0)=1$ and $v_{2}(0)=0$.

We are interested in the appearance of right polarized particles after
neutrinos interact with external fields. It is the situation, which
is implemented in neutrino spin oscillations: one looks for right
polarized neutrinos in a beam initially consisting of left particles
of the same type. Using Eq.~(\ref{eq:vsol}), one obtains that the
probability for $L\to R$ transitions has the form,
\begin{equation}\label{eq:PLRlambda}
  P_{\mathrm{L}\to\mathrm{R}}=
  \frac{1}{2}\tilde{\psi}^{\dagger}(1+\Sigma_{z})\tilde{\psi}=
  |v_{2}|^{2}=\frac{\mu^{2}B_{0}^{2}}{\Omega^{2}}\sin^{2}\Omega u_{0}.
\end{equation}
In general situation, $P_{\mathrm{L}\to\mathrm{R}}$ in Eq.~(\ref{eq:PLRlambda})
depends on $u_{0}=t-z$. We also note that this expression contains
the dependence on the quantum number $\lambda$ which does not have
a clear physical meaning yet. Hence, one should express $\lambda$
in terms of the neutrino energy and momentum.

The Hamiltonian of Eq.~(\ref{eq:DirEqHam}) explicitly depends on
$t$ and $z$. Thus the neutrino energy $E$ and the momentum $p_{z}$
along the wave propagation direction are not defined. Nevertheless
we can define the effective $E$ and $p_{z}$ as
\begin{align}
  E-p_{z} & =
  2\mathrm{i}\tilde{\psi}^{\dagger}\partial_{3}\tilde{\psi}=
  \lambda+\frac{V}{2},
  \label{eq:E-pzeff}
  \\
  E+p_{z} & =2\mathrm{i}\tilde{\psi}^{\dagger}\partial_{0}\tilde{\psi}=
  \lambda\frac{m^{2}}{\lambda^{2}-V^{2}/4}+\frac{V}{2}+
  2v^{\dagger}(\bm{\sigma}^{*}\mathbf{R})v.
  \label{eq:E+pzeff}
\end{align}
Using Eq.~(\ref{eq:vsol}), one gets that
\begin{equation}\label{eq:sigmaR}
  v^{\dagger}(\bm{\sigma}^{*}\mathbf{R})v=\pm
  \left(
    R_{z}+\frac{\rho}{2}
  \right),
  \quad
  \rho(\omega,u_{0})=2g\omega
  \frac{\mu^{2}B_{0}^{2}}{\Omega^{2}}\sin^{2}\Omega u_{0},
\end{equation}
where the signs $\pm$ stay for initially left and right polarized
neutrinos. A right polarized neutrino corresponds to $v_{0}^{\mathrm{T}}=(0,1)$
in Eq.~(\ref{eq:vsol}).

Finally, using Eqs.~(\ref{eq:E-pzeff})-(\ref{eq:sigmaR}), we have
the energy of left neutrinos as
\begin{equation}\label{eq:EL}
  E_{\mathrm{L}}=\frac{V+\rho}{2}+\sqrt{m^{2}+
  \left(
    p_{z}+\frac{V-\rho}{2}
  \right)^{2}} \approx
  p_{z}+\frac{m^{2}}{2p_{z}}+V,
\end{equation}
and
\begin{equation}\label{eq:ER}
  E_{\mathrm{R}}=\frac{V-\rho}{2}+\sqrt{m^{2}+
  \left(
    p_{z}-\frac{V-\rho}{2}
  \right)^{2}}\approx
  p_{z}+\frac{m^{2}}{2p_{z}},
\end{equation}
for right particles. The expansions in Eqs.~(\ref{eq:EL}) and~(\ref{eq:ER})
correspond to ultrarelativistic neutrinos with $p_{z}\gg m$. One
can see that, in this case, the energies have the conventional form.
However, for nonrelativistic neutrinos, the effective energies become
time and $z$ dependent.

The transition probability depends on the quantity $R_{z}$ given
in Eq.~(\ref{eq:R}). Using Eqs.~(\ref{eq:E-pzeff}) and~(\ref{eq:EL}),
one gets that $R_{z}$ has the following form for left neutrinos:
\begin{equation}\label{eq:Rz}
  R_{z}=-\frac{V}{4}
  \left(
    1-\frac{m^{2}}{\lambda^{2}-V^{2}/4}
  \right) \approx
  \frac{V}{2}
  \left(
    \frac{V}{p_{z}}+\frac{m^{2}}{2p_{z}^{2}}
  \right)^{-1}.
\end{equation}
Basing on Eqs.~(\ref{eq:PLRlambda}) and~(\ref{eq:Rz}), the transition
probability can be rewritten as
\begin{align}\label{eq:PLRpz}
  P_{\mathrm{L}\to\mathrm{R}} & =
  \frac{\mu^{2}B_{0}^{2}}{\Omega^{2}}\sin^{2}
  \left[
    \Omega(t-z)
  \right],
  \nonumber
  \\
  \Omega & \approx
  \sqrt{\mu^{2}B_{0}^{2}+
  \left[
    \frac{V}{2}
    \left(
      \frac{V}{p_{z}}+\frac{m^{2}}{2p_{z}^{2}}
    \right)^{-1}+
    \frac{g\omega}{2}
  \right]^{2}},
\end{align}
where we explicitly show the dependence on $t$ and $z$.

Now let us consider the quasiclassical approximation. In this limit,
a neutrino moves along a trajectory, which is a straight line $z=\beta t$,
where $\beta=p_{z}/E$ is the neutrino velocity. We can represent
the transition probability in Eq.~(\ref{eq:PLRpz}) in the following
way:
\begin{align}\label{eq:PLRqc}
  P_{\mathrm{L}\to\mathrm{R}}(t)= &
  \frac{\mu^{2}B_{0}^{2}(1-\beta)^{2}}{\mu^{2}B_{0}^{2}(1-\beta)^{2}+
  \left[
    \frac{V}{2}+\frac{g\omega}{2}(1-\beta)
  \right]^{2}}
  \nonumber
  \\
  & \times
  \sin^{2}
  \left(
    \sqrt{\mu^{2}B_{0}^{2}(1-\beta)^{2}+
    \left[
      \frac{V}{2}+\frac{g\omega}{2}(1-\beta)
    \right]^{2}}t
  \right),
\end{align}
since
\begin{equation}\label{eq:Vlim}
  \frac{V}{2}
  \left(
    \frac{V}{p_{z}}+\frac{m^{2}}{2p_{z}^{2}}
  \right)^{-1}
  (1-\beta) \approx\frac{V}{2}.
\end{equation}
The expression for $P_{\mathrm{L}\to\mathrm{R}}$ in Eq.~(\ref{eq:PLRqc})
coincides with the result of Ref.~\cite{EgoLobStu00}, where the
neutrino spin evolution in matter under the influence of a plane electromagnetic
wave was treated within the quasiclassical approach from the very
beginning.

\section{Spin-flavor oscillations: Great diagonal magnetic moments\label{sec:DIAG}}

Now we turn to the study of neutrino spin-flavor oscillations. Here
we are interested in the situation of great diagonal magnetic moments.
Basing on the results of Sec.~\ref{sec:SPINOSC}, we derive the analytical
transition probability for this type of oscillations.

Using Eq.~(\ref{eq:Lagrpsi}), we obtain the system of coupled Dirac
equations for the neutrino mass eigenstates $\psi_{a}$, $a=1,2$,
\begin{align}\label{eq:Direqsys}
  \mathrm{i}\partial_{t}\psi_{a} & =\mathcal{H}_{a}\psi_{a}+\mathcal{V}\psi_{b},
  \quad
  a\neq b,
  \nonumber
  \\
  \mathcal{H}_{a} & =-\mathrm{i}(\bm{\alpha}\nabla)+\beta m_{a}+
  \mu_{a}(\mathrm{i}\bm{\gamma}\mathbf{E}-\beta\bm{\Sigma}\mathbf{B})+
  \frac{V_{a}}{2}(1-\gamma^{5}),
  \nonumber
  \\
  \mathcal{V} & =\mu(\mathrm{i}\bm{\gamma}\mathbf{E}-\beta\bm{\Sigma}\mathbf{B})+
  \frac{V}{2}(1-\gamma^{5}),
\end{align}
where $V_{a}\equiv V_{aa}$ and $\mu_{a}\equiv\mu_{aa}$ for $a=1,2$,
$V\equiv V_{12}$, and $\mu\equiv\mu_{12}$ is the transition magnetic
moment.

We shall analyze the system in Eq.~(\ref{eq:Direqsys}) in the approximation
when $\mu_{a}\gg\mu$. There are multiple models of the neutrino magnetic
moments generation. Some of the models predict the diagonal magnetic
moments $\mu_{a}$ proportional to the neutrino masses $m_{a}$. In
these cases, the value of $\mu$ is suppressed because of the Glashow-Iliopoulos-Maiani
(GIM) mechanism~\cite{FukYan03}.

Moreover, we suppose that $|V_{a}|\gg V$. If we study $\nu_{e}\to\nu_{\mu,\tau}$
oscillations, then, using Eqs.~(\ref{eq:fnu}) and~(\ref{eq:nupsi}),
we get that
\begin{align}\label{eq:V12V}
  V_{1} & =\sqrt{2}G_{\mathrm{F}}
  \left(
    n_{e}\sin^{2}\theta-\frac{1}{2}n_{n}
  \right),
  \quad
  V_{2}=\sqrt{2}G_{\mathrm{F}}
  \left(
    n_{e}\cos^{2}\theta-\frac{1}{2}n_{n}
  \right),
  \nonumber
  \\
  V & =\frac{G_{\mathrm{F}}}{\sqrt{2}}n_{e}\sin2\theta.
\end{align}
Basing on Eq.~(\ref{eq:V12V}), one gets that the condition $|V_{a}|\gg V$
is satisfied if either $n_{n}\gg n_{e}$ or $\theta\ll1$. The former
case is implemented in a neutron rich environment like a neutron star.
The latter situation takes place if we study $\nu_{e}\to\nu_{\tau}$
oscillations since, as found in Ref.~\cite{An17}, $\theta\equiv\theta_{13}=0.15$
is much less than both $\theta_{\odot}=0.6$~\cite{Ago18} and $\theta_{\mathrm{ATM}}=(0.75\div0.85)$~\cite{Ace18}.

We are interested in spin-flavor oscillations of the type $\nu_{\beta\mathrm{L}}\to\nu_{\alpha\mathrm{R}}$,
i.e. when both flavor and the polarization are changed. If the above
approximations are satisfied, we can derive the analytical expression
for the transition probability for $\nu_{\beta\mathrm{L}}\to\nu_{\alpha\mathrm{R}}$
oscillations. Indeed, if we neglect $\mathcal{V}$ in Eq.~(\ref{eq:Direqsys}),
the neutrino spin evolves independently within each mass eigenstate,
as described in Sec.~\ref{sec:SPINOSC}. The transitions between
different neutrino flavors are solely owing to the vacuum neutrino
mixing. As in Sec.~\ref{sec:SPINOSC}, here we consider a neutrino
beam propagating along the electromagnetic wave.

To describe the evolution of neutrinos, we use the approach developed
in Ref.~\cite{Dvo11}, where the initial condition problem is solved.
The initial conditions corresponding to $\nu_{\beta\mathrm{L}}\to\nu_{\alpha\mathrm{R}}$
are the following. Since there are no right polarized neutrinos initially,
we choose $\nu_{\alpha\mathrm{R}}(z,0)=\nu_{\beta\mathrm{R}}(z,0)=0$.
The wave functions of left polarized neutrinos should be chosen like
$\nu_{\alpha\mathrm{L}}(z,0)=0$ and $\nu_{\beta\mathrm{L}}(z,0)\sim\exp(\mathrm{i}p_{z}z)$. Such a choice of the initial condition for $\nu_{\beta\mathrm{L}}(z,0)$
corresponds to a broad wave packet. The arbitrary initial wave packets
are discussed in Ref.~\cite{Dvo18}. Here the spin projections are
defined using the operators $(1\pm\Sigma_{z})/2$.

The projected wave functions of mass eigenstates, given in Eq.~(\ref{eq:psiproj}),
which satisfy the system in Eq.~(\ref{eq:Direqsys}), have the form,
\begin{equation}\label{eq:psias}
  \tilde{\psi}_{a}(z,t) =
  \sum_{s=\mathrm{L},\mathrm{R}}
  \exp
  \left(
    -\mathrm{i}E_{as}t+\mathrm{i}p_{z}z
  \right)
  a_{as}\tilde{u}_{as},
  \quad
  \tilde{u}_{as}^{\mathrm{T}}=
  \left(
    0,v_{s1}^{(a)},v_{s2}^{(a)},0
  \right),
\end{equation}
where the index $s=\mathrm{L},\mathrm{R}$ corresponds to initially
left or right polarized neutrinos and the energies $E_{as}$ are given
by Eqs.~(\ref{eq:EL}) and~(\ref{eq:ER}) with the replacements
$m\to m_{a}$ and $V\to V_{a}$. Since we neglect $\mathcal{V}$ in
Eq.~(\ref{eq:Direqsys}), the coefficients $a_{as}$ are constant
and entirely fixed by the initial condition. Using Eq.~(\ref{eq:nupsi}),
we get that $a_{1\mathrm{L}}=\sin\theta$ and $a_{2\mathrm{L}}=\cos\theta$.
Moreover $a_{1,2\mathrm{R}}=0$ since there are no right polarized
particles initially.

To describe the evolution of the spinors $v^{(a)}_{s}$ in Eq.~(\ref{eq:psias}),
we use the quasiclassical approximation from the very beginning, i.e.
we suppose that $z=\bar{\beta}t$, where $\bar{\beta}=2p_{z}/(E_{1\mathrm{L}}+E_{2\mathrm{L}})$
is the center of inertia velocity. Using Eq.~(\ref{eq:vsol}), we
get that the components of $v_{a\mathrm{L}}$ evolve as 
\begin{align}\label{eq:vaL}
  v_{\mathrm{L}1}^{(a)}(t) & =
  \exp
  \left[
    \mathrm{i}g\omega(1-\bar{\beta})t/2
  \right]
  \left\{
    \cos
    \left[
      \Omega_{a}(1-\bar{\beta})t
    \right]-
    \mathrm{i}n^{(a)}_{z}\sin
    \left[
      \Omega_{a}(1-\bar{\beta})t
    \right]
  \right\},
  \nonumber
  \\
  v_{\mathrm{L}2}^{(a)}(t) & = -
  \mathrm{i}n^{(a)}_{x}
  \exp
  \left[
    -\mathrm{i}g\omega(1-\bar{\beta})t/2
  \right]
  \sin
  \left[
    \Omega_{a}(1-\bar{\beta})t
  \right],
\end{align}
where the quantities $\Omega_{a}$ and $n^{(a)}_{x,z}$ are the natural
generalizations of the corresponding parameters given in Eqs.~(\ref{eq:UzOmega})
and~(\ref{eq:n}) with the replacements $m\to m_{a}$, $\mu\to\mu_{a}$,
and $V\to V_{a}$ there. The evolution of $v^{(a)}_{\mathrm{R}}$ is not
important since there are not right polarized neutrinos initially.

Basing on Eqs.~(\ref{eq:nupsi}), (\ref{eq:psias}), and~(\ref{eq:vaL}),
we derive the right polarized wave function $\nu_{\alpha\mathrm{R}}(z,t)=(1+\Sigma_{z})\left[\cos\theta\psi_{1}(z,t)-\sin\theta\psi_{2}(z,t)\right]/2$
in the following form:
\begin{align}\label{eq:nualphaR}
  \nu_{\alpha\mathrm{R}}^{\mathrm{T}}(z,t)= &
  \sin\theta\cos\theta
  \exp
  \left(
     \mathrm{i}p_{z}z
  \right)
  \nonumber
  \\
  & \times
  \left(
    0,0,
    \exp
    \left(
      -\mathrm{i}E_{1\mathrm{L}}t
    \right)
    v_{\mathrm{L}2}^{(1)}(t) -
    \exp
    \left(
      -\mathrm{i}E_{2\mathrm{L}}t
    \right)
    v_{\mathrm{L}2}^{(2)}(t),0
  \right).
\end{align}
Now, using Eq.~(\ref{eq:nualphaR}), one obtains the probability
for transitions $\nu_{\beta\mathrm{L}}\to\nu_{\alpha\mathrm{R}}$
in the form,
\begin{equation}\label{eq:PLRbigmua}
  P_{\beta\mathrm{L}\to\alpha\mathrm{R}}(t)=
  |\nu_{\alpha\mathrm{R}}(z,t)|^{2}=\sin^{2}(2\theta)
  \left[
    \frac{1}{4}(A_{1}-A_{2})^{2}+A_{1}A_{2}\sin^{2}(\Phi t)
  \right],
\end{equation}
where
\begin{align}\label{eq:Aa}
  A_{a}(t)= & \frac{\mu_{a}B_{0}(1-\bar{\beta})}
  {\sqrt{\mu_{a}^{2}B_{0}^{2}(1-\bar{\beta})^{2}+
  \left[
    \frac{V_{a}}{2}+\frac{g\omega}{2}(1-\bar{\beta})
  \right]^{2}}}
  \nonumber
  \\
  & \times
  \sin
  \left(
    \sqrt{\mu_{a}^{2}B_{0}^{2}(1-\bar{\beta})^{2}+
    \left[
      \frac{V_{a}}{2}+\frac{g\omega}{2}(1-\bar{\beta})
    \right]^{2}}t
  \right),
\end{align}
is the amplitude of spin oscillations within one mass eigenstate,
$\Phi=\Phi_{\mathrm{vac}}+(V_{1}-V_{2})/2$ is the phase of neutrino
flavor oscillations accounting for the matter contribution, $\Phi_{\mathrm{vac}}=\delta m^{2}/4p_{z}$
is the phase of neutrino oscillations in vacuum, and $\delta m^{2}=m_{1}^{2}-m_{2}^{2}$.
To derive Eq.~(\ref{eq:Aa}) we use the analogues of Eqs.~(\ref{eq:Rz})
and~(\ref{eq:Vlim}).

One can see that Eqs.~(\ref{eq:PLRbigmua}) and~(\ref{eq:Aa}) are
the generalization of the corresponding expressions obtained in Ref.~\cite{Dvo18}
for the situation when neutrinos interact not only with a plane electromagnetic
wave but also with the background matter. 

\section{Spin-flavor oscillations: Great transition magnetic moment\label{sec:TRANS}}

In this section, we continue to study spin-flavor oscillations. However,
unlike the case considered in Sec.~\ref{sec:DIAG}, we discuss the
situation of the great transition magnetic moment.

If $\mu\gg\mu_{a}$, we cannot neglect $\mathcal{V}$ in Eq.~(\ref{eq:Direqsys}).
It means that $a_{as}$ in Eq.~(\ref{eq:psias}) is no longer constant.
Analogously to Ref.~\cite{Dvo18} we suppose that $a_{as}=a_{as}(t-z)$.
Our main goal is to find the behavior of $a_{as}$. Moreover, in the
analogue of Eq.~(\ref{eq:psias}), we shall use the total wave function
$\psi_{as}$ rather than the projection $\tilde{\psi}_{as}$. Hence
we look for the solution of Eq.~(\ref{eq:Direqsys}) in the form,
\begin{align}\label{eq:psiastot}
  \psi_{a} & =
  \sum_{s=\mathrm{L},\mathrm{R}}
  \exp
  \left(
    -\mathrm{i}E_{as}t+\mathrm{i}p_{z}z
  \right)
  a_{as}(t-z)u_{as}.
\end{align}
%
We consider neutrinos
propagating along the wave in Eq.~(\ref{eq:psiastot}). Since $a_{as}$
is time dependent for both $s=\mathrm{L}$ and $s=\mathrm{R}$, we
should account for the time evolution of the basis spinors $u_{a\mathrm{L},\mathrm{R}}$.
For this purpose we choose two linearly independent initial spinors
$v_{0}$ for $a=1,2$, $v_{0\mathrm{L}}=(1,0)^{\mathrm{T}}$ and $v_{0\mathrm{R}}=(0,1)^{\mathrm{T}}$,
which contribute to Eq.~(\ref{eq:vsol}).

Substituting Eq.~(\ref{eq:psiastot}) to Eq.~(\ref{eq:Direqsys})
and taking into account that $\sim\exp\left(-\mathrm{i}E_{as}t+\mathrm{i}p_{z}z\right)u_{as}$
is the solution of the diagonal part of the system in Eq.~(\ref{eq:Direqsys}),
i.e. without $\mathcal{V}$, one gets the equation for the coefficients
$a_{as}$ in the form,
\begin{equation}
  \mathrm{i}\frac{1}{2}\sum_{s=\mathrm{L},\mathrm{R}}
  u_{as'}^{\dagger}(1-\alpha_{z})u_{as}\partial_{0}a_{as} =
  \frac{1}{2}
  \sum_{s=\mathrm{L},\mathrm{R}}
  u_{as'}^{\dagger}\mathcal{V}u_{bs}a_{bs}\exp[\mathrm{i}(E_{as'}-E_{bs})t].
\end{equation}
Using Eq.~(\ref{eq:basspinu}), we obtain the following mean values:
\begin{align}\label{eq:vavb}
  \frac{1}{2}u_{as'}^{\dagger}(1-\alpha_{z})u_{as} & =
  v_{as'}^{\dagger}v_{as}=v_{0s'}^{\dagger}v_{0s}=\delta_{ss'},
  \nonumber
  \\
  \frac{\mu}{2}
  u_{as'}^{\dagger}
  (\mathrm{i}\bm{\gamma}\mathbf{E}-\beta\bm{\Sigma}\mathbf{B})
  u_{bs} & =
  \mu v_{as'}^{\dagger}(\bm{\sigma}^{*}\mathbf{B})v_{bs},
  \nonumber
  \\
  \frac{V}{4}u_{as'}^{\dagger}(1-\gamma^{5})u_{bs} & =\frac{V}{2}
  \left(
    v_{as'2}^{*}v_{bs2}+
    \frac{m_{a}}{\lambda_{a}-V_{a}/2}
    \frac{m_{b}}{\lambda_{b}-V_{b}/2}v_{as'1}^{*}v_{bs1}
  \right).
\end{align}
Then we adopt the quasiclassical approximation, in which $\partial_{0}=(1-\bar{\beta})^{-1}\partial_{t}$,
where $\bar{\beta}$ is the mean velocity of the neutrino wave packet,
defined in Sec.~\ref{sec:DIAG}.

In this section, we consider the situation when $\mu_{a}\ll\mu$.
It means that the components of the vector $\mathbf{n}_{a}$, which
defines the neutrino spin evolution, have the following values: $n^{(a)}_{x}=0$
and $n^{(a)}_{z}=1$. We can use Eq.~(\ref{eq:vsol}) to compute the mean
values of the spinors $v_{as}$ in Eq.~(\ref{eq:vavb}) assuming
that the electromagnetic wave has the circular polarization. Then
we define the effective wave function $\Psi^{\mathrm{T}}=\left(a_{1\mathrm{R}},a_{1\mathrm{L}},a_{2\mathrm{R}},a_{2\mathrm{L}}\right)$,
which obeys the Schr\"odinger equation,
\begin{align}\label{eq:SchrodPsi}
  \mathrm{i}\frac{\mathrm{d}\Psi}{\mathrm{d}t} & = H\Psi,
  \nonumber
  \\
  H & =
  \left(
    \begin{array}{cccc}
      0 & 0 & 0 & \mu B_{0}(1-\bar{\beta})e^{\mathrm{i}\phi_{2}t}\\
      0 & 0 & \mu B_{0}(1-\bar{\beta})e^{\mathrm{i}\phi_{1}t} & Ve^{\mathrm{i}\phi't}\\
      0 & \mu B_{0}(1-\bar{\beta})e^{-\mathrm{i}\phi_{1}t} & 0 & 0\\
      \mu B_{0}(1-\bar{\beta})e^{-\mathrm{i}\phi_{2}t} & Ve^{-\mathrm{i}\phi't} & 0 & 0
    \end{array}
  \right).
\end{align}
Here
\begin{align}\label{eq:phi12}
  \phi_{2} & \approx\frac{\delta m^{2}}{2p_{z}}-V_{2}-g\omega(1-\bar{\beta}),
  \quad
  \phi_{1}\approx\frac{\delta m^{2}}{2p_{z}}+V_{1}+g\omega(1-\bar{\beta}),
  \nonumber
  \\
  \phi' & \approx\frac{\delta m^{2}}{2p_{z}}+V_{1}-V_{2},
\end{align}
where we take that neutrinos are ultrarelativistic particles.

Let us introduce the new wave function $\tilde{\Psi}^{\mathrm{T}}=\left(\tilde{a}_{1\mathrm{R}},\tilde{a}_{1\mathrm{L}},\tilde{a}_{2\mathrm{R}},\tilde{a}_{2\mathrm{L}}\right)$
as $\Psi=\mathcal{U}\tilde{\Psi}$, where 
\begin{align}\label{eq:calU}
  \mathcal{U}= & \text{diag}
  \bigg\{
    \exp
    \left[
      \mathrm{i}
      \left(
        \Phi_{\mathrm{-}}-\frac{V_{1}+V_{2}}{4}
      \right)
      t
    \right],
    \exp
    \left[
      \mathrm{i}
      \left(
        \Phi_{+}+\frac{3V_{1}-V_{2}}{4}
      \right)
      t
    \right],
    \nonumber
    \\
   & 
    \exp
    \left[
      -\mathrm{i}
      \left(
        \Phi_{\mathrm{+}}+\frac{V_{1}+V_{2}}{4}
      \right)
      t
    \right],
    \exp
    \left[
      -\mathrm{i}
      \left(
        \Phi_{\mathrm{-}}-\frac{3V_{2}-V_{1}}{4}
      \right)
      t
    \right]
  \bigg\} .
\end{align}
Here $\Phi_{\mathrm{\pm}}=\Phi_{\mathrm{vac}}\pm(1-\bar{\beta})g\omega/2$.
The wave function $\tilde{\Psi}$ obeys the equation
\begin{align}\label{eq:tildeH}
  \mathrm{i}\frac{\mathrm{d}\tilde{\Psi}}{\mathrm{d}t} = &
  \tilde{H}\tilde{\Psi},
  \nonumber
  \\
  \tilde{H}= &
  \mathcal{U}^{\dagger}H\mathcal{U}-i\mathcal{U}^{\dagger}\dot{\mathcal{U}}\nonumber \\
  & =
  \left(
    \begin{array}{cccc}
      \Phi_{\mathrm{-}}-\frac{V_{1}+V_{2}}{4} & 0 & 0 & \mu B_{0}(1-\bar{\beta})\\
      0 & \Phi_{\mathrm{+}}+\frac{3V_{1}-V_{2}}{4} & \mu B_{0}(1-\bar{\beta}) & V\\
      0 & \mu B_{0}(1-\bar{\beta}) & -\Phi_{\mathrm{+}}-\frac{V_{1}+V_{2}}{4} & 0\\
      \mu B_{0}(1-\bar{\beta}) & V & 0 & -\Phi_{\mathrm{-}}+\frac{3V_{2}-V_{1}}{4}
    \end{array}
  \right).
\end{align}
One can see that the effective Hamiltonian $\tilde{H}$ in Eq.~(\ref{eq:tildeH})
generalizes the analogous effective Hamiltonian derived in Ref.~\cite{Dvo18}
for the nonzero interaction of neutrinos with background matter. Moreover,
if set $(1-\bar{\beta})\to1$ and $\omega\to0$ in Eq.~(\ref{eq:tildeH}),
we reproduce the effective Hamiltonian for neutrino spin-flavor oscillations
in matter under the influence of a transverse magnetic field derived
in Ref.~\cite{Dvo12} using the relativistic quantum mechanics approach.

The solution of the Schr\"odinger equation in Eq.~(\ref{eq:tildeH})
results in the algebraic characteristic equation of the forth order,
which implies quite cumbersome expressions for eigenvalues and
eigenvectors. That is why we again suppose that $V\ll |V_{a}|$, as
in Sec.~\ref{sec:DIAG}, to proceed with the analytical solution.
The validity of this approximation will be discussed below.

In this case, the evolution of $\tilde{\Psi}$ can be represented
in the form,
\begin{equation}\label{eq:tildePsisol}
  \tilde{\Psi}(t)=\sum_{\substack{a=1,2 \\ \zeta=\pm}}
  \exp
  \left(
    -\mathrm{i}\mathcal{E}_{a}^{(\zeta)}t
  \right)
  \left(
    U_{a}^{(\zeta)}\otimes U_{a}^{(\zeta)\dagger}
  \right)
  \tilde{\Psi}(0),
\end{equation}
where
\begin{equation}\label{eq:Omega12}
  \mathcal{E}_{1,2}^{(\zeta)}=\pm\frac{V_{1}-V_{2}}{4}+\zeta\Omega_{1,2},
  \quad
  \Omega_{1,2}=\sqrt{(\mu B_{0})^{2}(1-\bar{\beta})^{2}+
  \left(
    \Phi_{\mathrm{\pm}}\pm\frac{V_{1,2}}{2}
  \right)^{2}},
\end{equation}
are the eigenvalues of the Hamiltonian $\tilde{H}$ in Eq.~(\ref{eq:tildeH}) and
\begin{align}\label{eq:Ubasspin}
  U_{1}^{+} & =\sqrt{\frac{\Omega_{1}+\Phi_{+}+V_{1}/2}{2\Omega_{1}}}
  \left(
    \begin{array}{c}
      0\\
      1\\
      \frac{\mu B_{0}(1-\bar{\beta})}{\Omega_{1}+\Phi_{+}+V_{1}/2}\\
      0
    \end{array}
  \right),
  \nonumber
  \\
  U_{1}^{-} & =\sqrt{\frac{\Omega_{1}+\Phi_{+}+V_{1}/2}{2\Omega_{1}}}
  \left(
    \begin{array}{c}
      0\\
      -\frac{\mu B_{0}(1-\bar{\beta})}{\Omega_{1}+\Phi_{+}+V_{1}/2}\\
      1\\
      0
    \end{array}
  \right),
  \nonumber
  \\
  U_{2}^{+} & =\sqrt{\frac{\Omega_{2}+\Phi_{-}-V_{2}/2}{2\Omega_{2}}}
  \left(
    \begin{array}{c}
      1\\
      0\\
      0\\
      \frac{\mu B_{0}(1-\bar{\beta})}{\Omega_{2}+\Phi_{-}-V_{2}/2}
    \end{array}
  \right),
  \nonumber
  \\
  U_{2}^{-} & =\sqrt{\frac{\Omega_{2}+\Phi_{-}-V_{2}/2}{2\Omega_{2}}}
  \left(
    \begin{array}{c}
      -\frac{\mu B_{0}(1-\bar{\beta})}{\mathcal{E}_{2}+\Phi_{-}-V_{2}/2}\\
      0\\
      0\\
      1
    \end{array}
  \right),
\end{align}
are the eigenvectors.

Equation~(\ref{eq:tildePsisol}) should be supplied with the initial condition
of the form,
\begin{equation}\label{eq:ictildePsi}
  \tilde{\Psi}^{\mathrm{T}}(0)=\left(0,\sin\theta,0,\cos\theta\right),
\end{equation}
which means that there are only neutrinos of the type $\nu_{\beta\mathrm{L}}$
initially. Using Eqs.~(\ref{eq:tildePsisol})-(\ref{eq:ictildePsi}),
one gets that the coefficients $a_{1,2\mathrm{R}}$ are expressed
in the following way: 
\begin{align}\label{eq:a12R}
  a_{1\mathrm{R}}(t) & =-\mathrm{i}\frac{\mu B_{0}}{\Omega_{2}}(1-\bar{\beta})
  \exp
  \left[
    \mathrm{i}
    \left(
      \Phi_{-}-\frac{V_{2}}{2}
    \right)
    t
  \right]
  \cos\theta\sin\Omega_{2}t,
  \nonumber
  \\
  a_{2\mathrm{R}}(t) & =-\mathrm{i}\frac{\mu B_{0}}{\Omega_{1}}(1-\bar{\beta})
  \exp
  \left[
    -\mathrm{i}
    \left(
      \Phi_{+}+\frac{V_{1}}{2}
    \right)
    t
  \right]
  \sin\theta\sin\Omega_{1}t.
\end{align}
The values of $a_{1,2\mathrm{L}}$ are not important for our purposes
since we are interested in spin-flavor oscillations when both flavor
and helicity change. 

Basing on Eqs.~(\ref{eq:nupsi}), (\ref{eq:psiastot}), and~(\ref{eq:a12R}),
the neutrino wave function $\nu_{\alpha\mathrm{R}}(z,t)=\cos\theta\psi_{1\mathrm{R}}(z,t)-\sin\theta\psi_{2\mathrm{R}}(z,t)$
reads
\begin{align}\label{eq:nuRfin}
  \nu_{\alpha\mathrm{R}}(z,t)= &
  \exp(\mathrm{i}p_{z}z)
  \left[
    \cos\theta\exp(-\mathrm{i}E_{1\mathrm{R}}t)a_{1\mathrm{R}}(t) -
    \sin\theta\exp(-\mathrm{i}E_{2\mathrm{R}}t)a_{2\mathrm{R}}(t)
  \right]
  \nu_{\mathrm{R}}
  \nonumber
  \\
  & =
  - \mathrm{i}\exp
  \left(
    -\mathrm{i}
    \left[
      p_{z}+\frac{m_{1}^{2}+m_{2}^{2}}{4p_{z}}+\frac{g\omega}{2}(1-\bar{\beta})
    \right] +
    \mathrm{i}p_{z}z
  \right)
  \mu B_{0}(1-\bar{\beta})
  \nonumber
  \\
  & \times
  \left[
    \cos^{2}\theta
    \exp
    \left(
      -\mathrm{i}\frac{V_{2}}{2}t
    \right)
    \frac{\sin\Omega_{2}t}{\Omega_{2}} -
    \sin^{2}\theta
    \exp
    \left(
      -\mathrm{i}\frac{V_{1}}{2}t
    \right)
    \frac{\sin\Omega_{1}t}{\Omega_{1}}
  \right]
  \nu_{\mathrm{R}},
\end{align}
where $\nu_{\mathrm{R}}$ is the constant bispinor satisfying $|\nu_{\mathrm{R}}|^{2}=1$
and $\Sigma_{z}\nu_{\mathrm{R}}=\nu_{\mathrm{R}}$.

The probability for transitions $\nu_{\beta\mathrm{L}}\to\nu_{\alpha\mathrm{R}}$
is derived using Eq.~(\ref{eq:nuRfin}) as
\begin{equation}\label{eq:PLRapprox}
  P_{\beta\mathrm{L}\to\alpha\mathrm{R}}(t)=
  |\nu_{\alpha\mathrm{R}}(z,t)|^{2}=
  \left\{
    \left[
      \mathcal{A}_{2}-\mathcal{A}_{1}
    \right]^{2} +
    4\mathcal{A}_{1}\mathcal{A}_{2}\sin^{2}
    \left(
      \frac{V_{1}-V_{2}}{4}t
    \right)
  \right\},
\end{equation}
where
\begin{equation}\label{eq:A12}
  \mathcal{A}_{1}=\mu B_{0}(1-\bar{\beta})\sin^{2}\theta
  \frac{\sin\Omega_{1}t}{\Omega_{1}},
  \quad
  \mathcal{A}_{2}=\mu B_{0}(1-\bar{\beta})\cos^{2}\theta
  \frac{\sin\Omega_{2}t}{\Omega_{2}},
\end{equation}
are the amplitudes of the transitions $\psi_{(1,\mathrm{L}),(2,\mathrm{R})}\leftrightarrow\psi_{(2,\mathrm{R}),(1,\mathrm{L})}$
in matter under the influence of an electromagnetic wave. The analogue
of $\mathcal{A}_{1,2}$ for the constant transverse magnetic fields
was introduced in Ref.~\cite{DvoMaa07}.

\begin{figure}
  \centering
  \subfigure[]
  {\label{1a}
  \includegraphics[scale=.35]{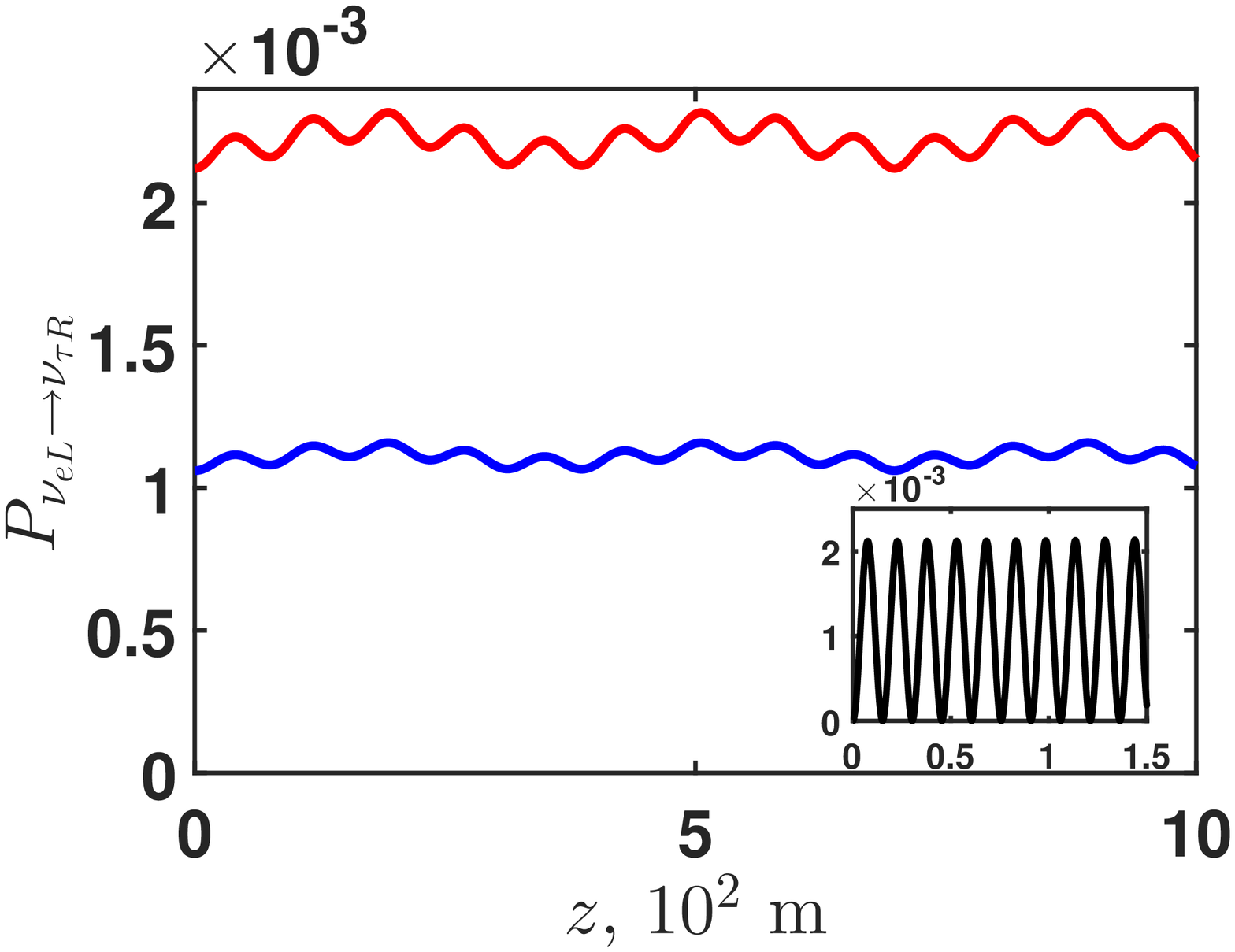}}
  \hskip-.6cm
  \subfigure[]
  {\label{1b}
  \includegraphics[scale=.35]{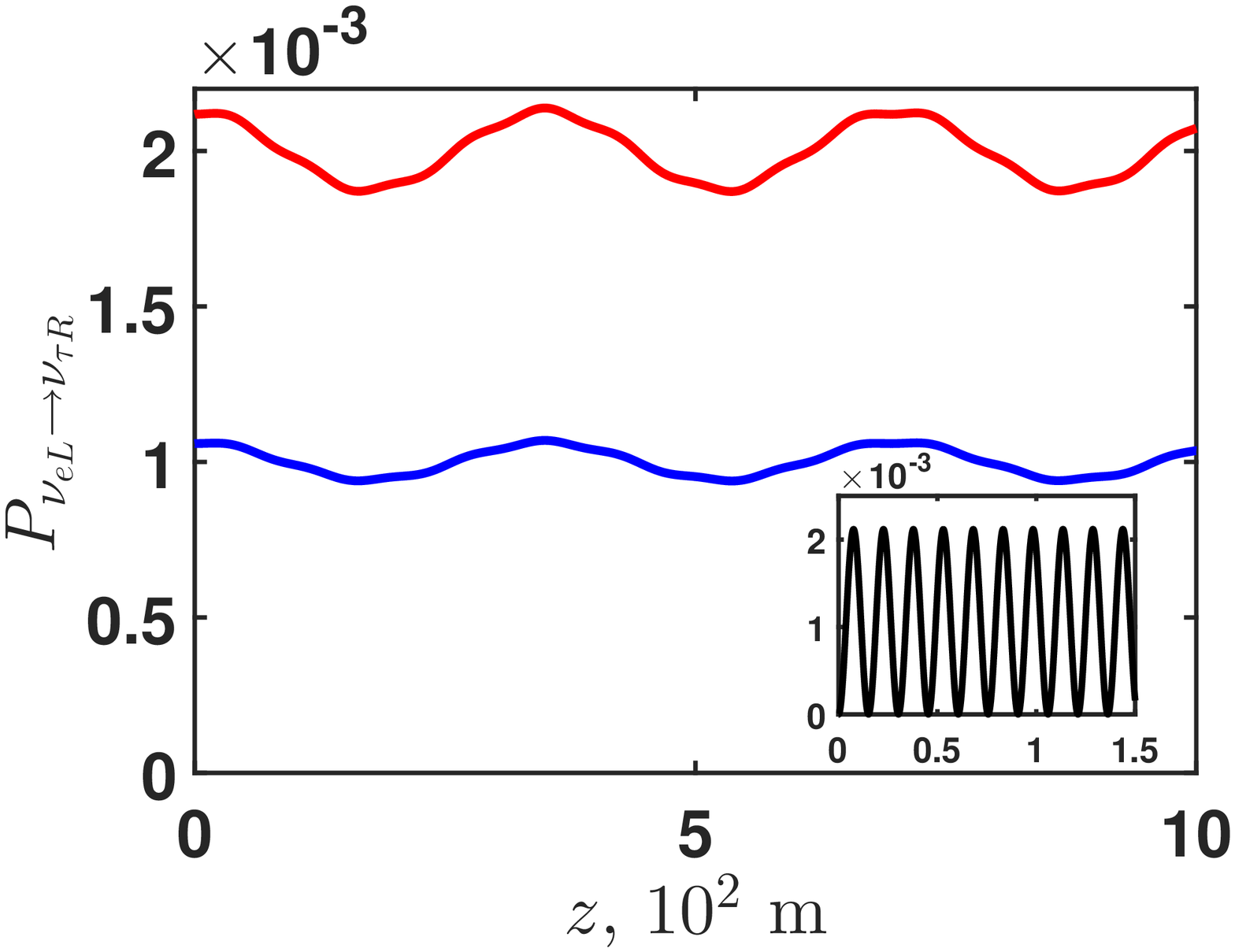}}
  \protect
  \caption{The transition probabilities for $\nu_{e\mathrm{L}}\to\nu_{\tau\mathrm{R}}$
  oscillations in the electroneutral hydrogen plasma with $n_{e}=10^{29}\,\text{cm}^{-3}$
  under the influence of the electromagnetic wave with $B_{0}=10^{18}\,\text{G}$
  and $\omega=10^{13}\,\text{s}^{-1}$ versus the distance $z=\bar{\beta}t$
  traveled by the neutrino beam. The parameters of neutrinos are
  $\delta m^{2}=2.5\times10^{-3}\,\text{eV}^{2}$~\cite{Est18},
  $\theta=0.15$~\cite{An17}, $p_{z}=1\,\text{keV}$,
  and $\mu=10^{-11}\mu_{\mathrm{B}}$~\cite{Bed13}.
  (a) The approximate transition probability in Eq.~(\ref{eq:PLRapprox})
  corresponding to the case $V=0$ in $\tilde{H}$ in Eq.~(\ref{eq:tildeH}).
  (b) The transition probability in Eq.~(\ref{eq:PLRnumsol}) based
  on the numerical solution of Eq.~(\ref{eq:tildeH}) with $V\protect\neq0$.
  Red and blue lines are the upper envelope function and the averaged
  transition probability. The insets in panels~(a) and~(b) show
  $P_{\nu_{e\mathrm{L}}\to\nu_{\tau\mathrm{R}}}(z)$
  at $0<z<150\,\text{m}$.\label{fig:eLtauR}}
\end{figure}

The behavior of the transition probability in Eq.~(\ref{eq:PLRapprox})
is shown in Fig.~\ref{1a} for $\nu_{e\mathrm{L}}\to\nu_{\tau\mathrm{R}}$
oscillations channel versus the distance $z\approx t$ passed by the
neutrino beam. We suppose that the electromagnetic wave has the following
characteristics: $B_{0}=10^{18}\,\text{G}$ and $\omega=10^{13}\,\text{s}^{-1}$.
The neutrino energy and the transition magnetic moment are taken to
be $E_{\nu}\equiv p_{z}=1\,\text{keV}$ and $\mu=10^{-11}\mu_{\mathrm{B}}$, where $\mu_{\mathrm{B}}$ is the Bohr magneton.
As mentioned in Ref.~\cite{Dvo18}, these parameters can model neutrino
spin-flavor oscillations in the vicinity of a highly magnetized pulsar.
To estimate the mean velocity of neutrinos $\bar{\beta}$ we assume
that the neutrino masses are on the level of $1\,\text{eV}$~\cite{Ase11}.

The motivation for the choice of the matter density value in Fig.~\ref{fig:eLtauR}
is the following. We can consider neutrino spin-flavor oscillations
in the vicinity of a compact astrophysical object surrounded by an
accretion disk. For example, properties of a gamma-ray burst (GRB)
can be explained by the matter accretion to a central object. In this
model of GRB, the matter density of a hydrogen plasma in the inner
part of an accretion disk can reach $10^{26}\,\text{cm}^{-3}$~\cite{PopWooFry99}
or be even higher~\cite{MatPerNar}. Such values of $n_{e}$ are
close to these used in our simulations (especially see Fig.~\ref{fig:eLmuR}
below). Note that this model of GRB predicts a high neutrino emissivity
by an accretion disk~\cite{PopWooFry99,MatPerNar}.%

The function $P_{\nu_{e\mathrm{L}}\to\nu_{\tau\mathrm{R}}}(z)$ is
a rapidly oscillating one. It is the typical feature of a neutrino
system which experiences spin-flavor oscillations in matter and an
electromagnetic field with different oscillations frequencies induced
by matter and an electromagnetic field; cf. Refs.~\cite{Dvo18,Dvo11}.
That is why, here, we show only the upper envelope function and the
averaged transition probability. The upper envelope function is built
using the spline interpolation of the maxima of $P_{\nu_{e\mathrm{L}}\to\nu_{\tau\mathrm{R}}}(z)$.
The transition probability $P_{\nu_{e\mathrm{L}}\to\nu_{\tau\mathrm{R}}}(z)$
is shown only in the inset in Fig.~\ref{1a} for small
$z$.

One can see in Fig.~\ref{1a} that the transition probability
for the considered oscillations channel reaches only a tiny value
$\sim10^{-3}$. This fact can be explained by the great value of $\Phi_{\mathrm{vac}}$
for $\nu_{e\mathrm{L}}\to\nu_{\tau\mathrm{R}}$ oscillations, which
is about 2 orders of magnitude greater than other entries in $\tilde{H}$
in Eq.~\eqref{eq:tildeH}. Hence $\Omega_{1,2}\gg\mu B_{0}(1-\bar{\beta})$
and $\mathcal{A}_{1,2}\ll1$ in Eq.~(\ref{eq:A12}).

Now we compare the exact solution, given in Eqs.~(\ref{eq:PLRapprox})
and~(\ref{eq:A12}), of the approximate effective Schr\"odinger Eq.~(\ref{eq:tildeH}),
where we put $V=0$, with the numerical solution of the exact Eq.~(\ref{eq:tildeH}).
Should one have the solution $\tilde{\Psi}^{\mathrm{T}}(t)=(\tilde{\Psi}_{1},\tilde{\Psi}_{2},\tilde{\Psi}_{3},\tilde{\Psi}_{4})$
of Eq.~(\ref{eq:tildeH}), supplied with the initial condition in
Eq.~(\ref{eq:ictildePsi}), the transition probability for $\nu_{\beta\mathrm{L}}\to\nu_{\alpha\mathrm{R}}$
oscillations can be found as 
\begin{equation}\label{eq:PLRnumsol}
  P_{\beta\mathrm{L}\to\alpha\mathrm{R}}(t) =
  \left|
    \cos\theta\tilde{\Psi}_{1}(t)-\sin\theta\tilde{\Psi}_{3}(t)
  \right|^{2}.
\end{equation}
Equation~(\ref{eq:PLRnumsol}) can be verified with help of Eqs.~(\ref{eq:calU})
and~(\ref{eq:nuRfin}).

In Fig.~\ref{1b}, we show the transition probability
for $\nu_{e\mathrm{L}}\to\nu_{\tau\mathrm{R}}$ oscillations based
on Eq.~(\ref{eq:PLRnumsol}) calculated using the numerical solution
of Eq.~(\ref{eq:tildeH}) with $V\neq0$. The transition probability
$P_{\nu_{e\mathrm{L}}\to\nu_{\tau\mathrm{R}}}(z)$ corresponds to
the same parameters of the neutrino system and the external fields,
which are used in Fig.~\ref{1a}. Comparing Figs.~\ref{1a}
and~\ref{1b}, one can see that the upper envelope function,
depicted by the red line, and the averaged transition probability,
shown by the blue line, oscillate near the mean values $\approx2\times10^{-3}$
and $\approx10^{-3}$ respectively. Despite the frequencies of this
oscillation are different, the mean values of the upper envelope function
and the averaged transition probability are practically the same.
Thus the exact solution in Eqs.~(\ref{eq:PLRapprox}) and~(\ref{eq:A12})
of the approximate Schr\"odinger Eq.~(\ref{eq:tildeH}) with $V=0$
represents a qualitatively correct description of $\nu_{e\mathrm{L}}\to\nu_{\tau\mathrm{R}}$
oscillations.

Now we consider $\nu_{e\mathrm{L}}\to\nu_{\mu\mathrm{R}}$ oscillations
channel. In this situation, we cannot neglect $V$ in Eq.~(\ref{eq:tildeH})
since $\theta\equiv\theta_{\odot}=0.6$ is not small. That is why
Eqs.~(\ref{eq:PLRapprox}) and~(\ref{eq:A12}) are not applicable
and we have to use the numerical solution of Eq.~(\ref{eq:tildeH})
from the very beginning.

In Fig.~\ref{2a}, we show the transition probability $P_{\nu_{e\mathrm{L}}\to\nu_{\mu\mathrm{R}}}(z)$,
the upper and lower envelope functions, and the averaged transition
probability. The values of the parameters of the external fields and
the neutrino system, except $\delta m^{2}$ and $\theta$, are the
same as in Fig.~\ref{fig:eLtauR}. One can see in Fig.~\ref{2a}
that the averaged transition probability oscillates near $5\%$ value.
It is much greater than in Fig.~\ref{1a}. This feature
can be explained by the fact that all the entries of $\tilde{H}$
in Eq.~(\ref{eq:tildeH}) are of the same order of magnitude for
$\nu_{e\mathrm{L}}\to\nu_{\mu\mathrm{R}}$ oscillations unlike the $\nu_{e\mathrm{L}}\to\nu_{\tau\mathrm{R}}$
channel, in which $\Phi_{\mathrm{vac}}$ is dominant.

\begin{figure}
  \centering
  \subfigure[]
  {\label{2a}
  \includegraphics[scale=.35]{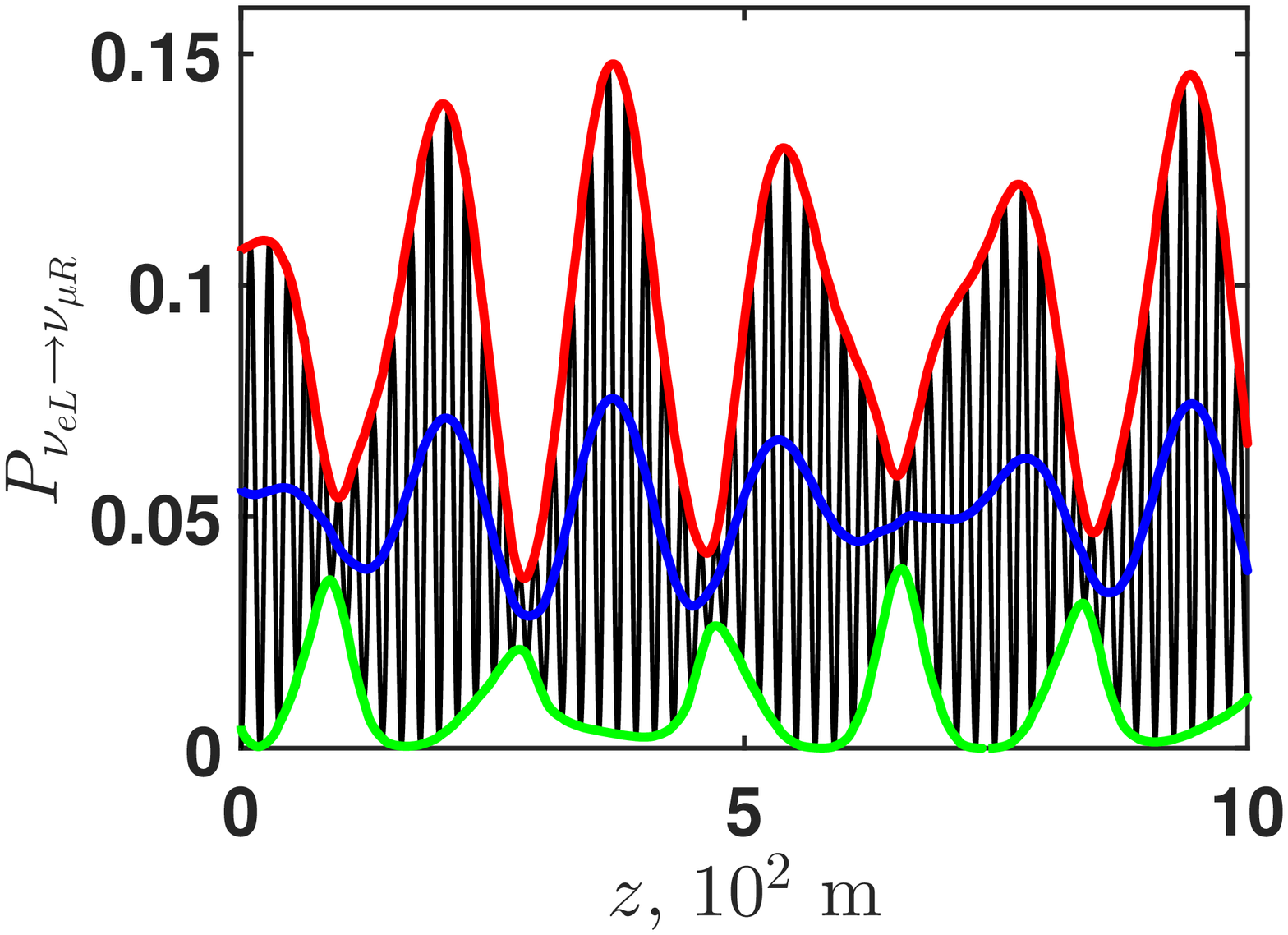}}
  \hskip-.6cm
  \subfigure[]
  {\label{2b}
  \includegraphics[scale=.35]{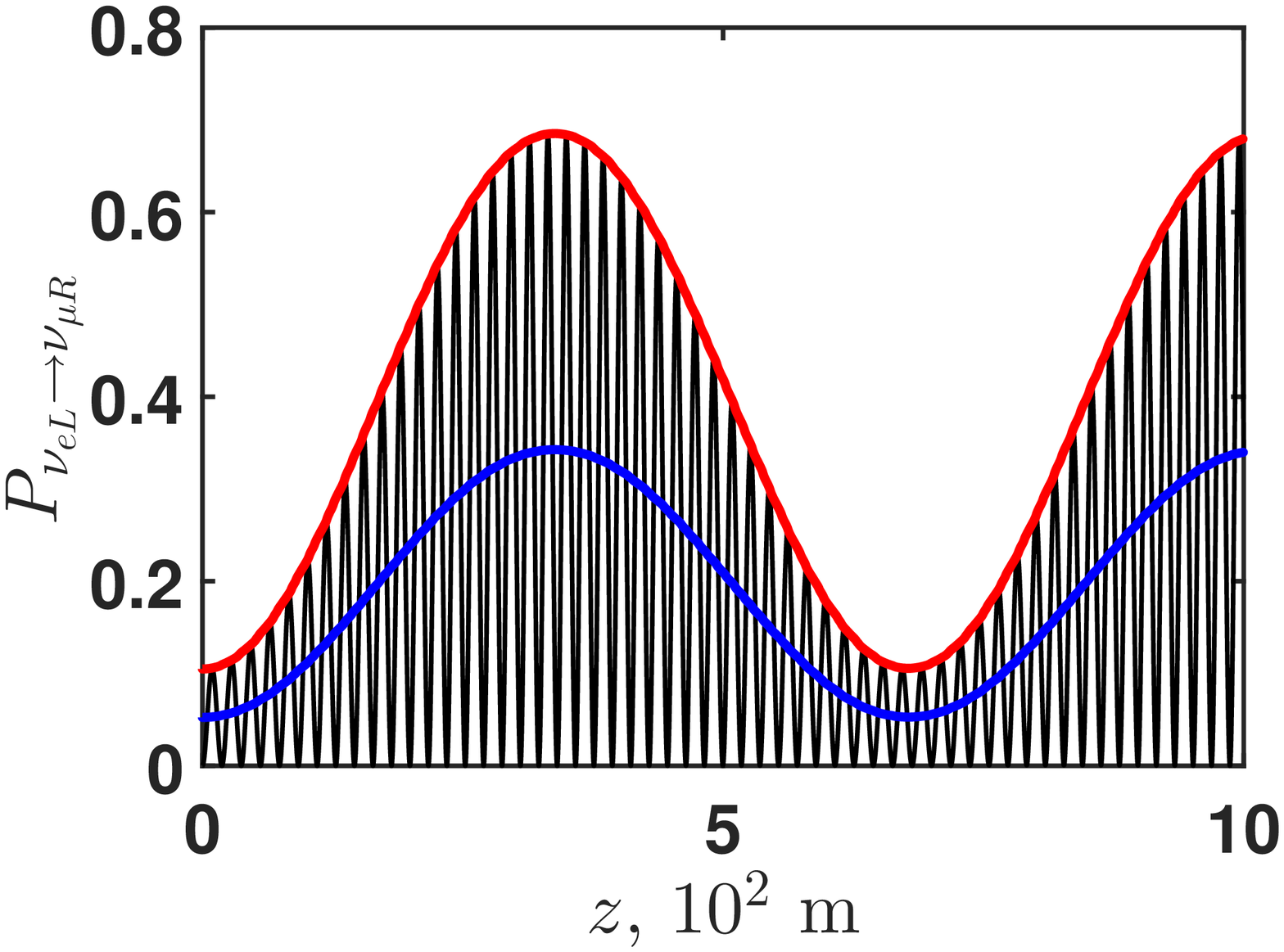}}
  \protect
  \caption{The transition probabilities for $\nu_{e\mathrm{L}}\to\nu_{\mu\mathrm{R}}$
  oscillations in the electroneutral hydrogen plasma when particles
  interact with the electromagnetic wave having $B_{0}=10^{18}\,\text{G}$
  and $\omega=10^{13}\,\text{s}^{-1}$ versus the distance $z=\bar{\beta}t$
  passed by the neutrino beam. The parameters of neutrinos are
  $\delta m^{2}=7.59\times10^{-5}\,\text{eV}^{2}$~\cite{Abe16},
  $\theta=0.6$~\cite{Ago18}, $p_z=1\,\text{keV}$, and $\mu=10^{-11}\mu_{\mathrm{B}}$.
  These transition probabilities correspond to Eq.~(\ref{eq:PLRnumsol}),
  which is based on the numerical solution of Eq.~(\ref{eq:tildeH})
  with $V\protect\neq0$. (a) $n_{e}=10^{29}\,\text{cm}^{-3}$; and 
  (b) $n_{e}=10^{27}\,\text{cm}^{-3}$. Red and blue lines are the upper
  envelope functions and the averaged transition probabilities. The
  green line in panel~(a) is the lower envelope function.\label{fig:eLmuR}}
\end{figure}

In Fig.~\ref{2b}, we depict $P_{\nu_{e\mathrm{L}}\to\nu_{\mu\mathrm{R}}}(z)$
for lower matter density $n_{e}=10^{27}\,\text{cm}^{-3}$, which is
very close to the value in the inner part of an accretion disk predicted
by the model of GRB in Ref.~\cite{MatPerNar}. The transition probability
in this case reproduces the result in Ref.~\cite{Dvo18}, where spin-flavor
oscillations $\nu_{e\mathrm{L}}\to\nu_{\mu\mathrm{R}}$ were described
at the absence of the matter contribution. Comparing Figs.~\ref{2a}
and~\ref{2b}, one can see that the lower matter density
is, the higher transition probability is. Thus, one does not expect
the appearance of a resonance in neutrino spin-flavor oscillations
in matter under the influence of a plane electromagnetic wave, as
claimed in Ref.~\cite{EgoLobStu00}. The highest transition probability
can be observed when neutrinos do not interact with background matter.

To highlight the difference between our results and the findings of Ref.~\cite{EgoLobStu00} we present the transition probability for $\nu_{\beta\mathrm{L}}\to\nu_{\alpha\mathrm{R}}$, which can be derived on the basis of Eq.~(21) in Ref.~\cite{EgoLobStu00}. It has the form,
\begin{align}\label{eq:Pwrong}
  P_{\nu_{\beta\mathrm{L}}\to\nu_{\alpha\mathrm{R}}}(t) = &
  \frac{\mu^{2}B_{0}^{2}(1-\bar{\beta})^{2}}
  {\mu^{2}B_{0}^{2}(1-\bar{\beta})^{2} +
  \varDelta^2}
  \sin^2
  \left(
    \sqrt{\mu^{2}B_{0}^{2}(1-\bar{\beta})^{2} +
   \varDelta^2} t
  \right),
  \notag
  \\
  \varDelta = & \frac{\delta m^2}{4p_z} A(\theta) -  \frac{G_\mathrm{F}n_e}{\sqrt{2}} +   
  \frac{g\omega}{2}(1-\bar{\beta}),
\end{align}
where take into account that, for $\nu_{e\mathrm{L}}\to\nu_{\mu,\tau\mathrm{R}}$ oscillations channel, $A(\theta) =  (1+\cos2\theta)/2$~\cite{LikStu95} and $f_{\nu_e} - f_{\nu_{\mu,\tau}} = \sqrt{2} G_\mathrm{F}n_e$; cf Eq.~\eqref{eq:fnu}. 

One can see in Eq.~\eqref{eq:Pwrong} that the amplitude of the transition probability would become $\sim 1$ if $\varDelta = 0$. This fact contradicts to out results both in Eqs.~\eqref{eq:PLRapprox} and~\eqref{eq:A12} and the numerical simulations shown in Figs.~\ref{fig:eLtauR} and~\ref{fig:eLmuR}. This inconsistency can be accounted for by the incorrect generalization of the Bargmann-Michel-Telegdi equation for the description of neutrino spin-flavor oscillations. In general situation, when one studies spin-flavor oscillations of Dirac neutrinos, an effective Schr\"odinger  equation cannot have a $2\times 2$ Hamiltonian. Typically, in this kind of problems, one deals with the system of $4$ differential equations, e.g., as in Eq.~\eqref{eq:SchrodPsi} or Eq.~\eqref{eq:tildeH}.

\section{Conclusion}\label{sec:CONCL}

In the present work, we have studied neutrino spin and spin-flavor oscillations
in matter under the influence of a plane electromagnetic wave with
the circular polarization. Neutrinos are supposed to be massive Dirac
particles with nonzero mixing between different neutrino flavors,
and possessing arbitrary matrix of magnetic moments. We have started
in Sec.~\ref{sec:INTERACT} with reminding the basic features of
neutrino interaction with background matter and an electromagnetic
field.

In Sec.~\ref{sec:SOLDIREQ}, we have found the new exact solution
of the Dirac-Pauli equation for a massive neutrino with a nonzero
magnetic moment interacting with matter under the influence a plane
electromagnetic wave. Previously, the solution of the wave equation
for a Dirac fermion with an anomalous magnetic moment interacting
with a plane electromagnetic wave in vacuum, i.e. at the absence of
the electroweak background matter, was known (see, e.g., Ref.~\cite{BagGit90}).

In Sec.~\ref{sec:SPINOSC}, we have applied the solution obtained
in Sec.~\ref{sec:SOLDIREQ} for the description of neutrino spin
oscillation in the considered external fields. We have studied the
process $\nu_{\mathrm{L}}\to\nu_{\mathrm{R}}$, that is the neutrino
spin precession within one neutrino mass eigenstate. The probability
$P_{\mathrm{L}\to\mathrm{R}}$ for transitions of this kind has been
derived. We have demonstrated that, in the quasiclassical approximation,
the expression for $P_{\mathrm{L}\to\mathrm{R}}$ in Eq.~(\ref{eq:PLRqc})
coincides with the result of Ref.~\cite{EgoLobStu00}, where the
neutrino spin evolution in the external fields was studied within
the quasiclassical approach from the very beginning.

Then, we have turned to the consideration of spin-flavor oscillations.
For this purpose we have formulated the initial condition problem.
This approach for the description of neutrino flavor and spin-flavor
oscillations in constant external fields has been developed in Ref.~\cite{Dvo11}
earlier.

First, in Sec.~\ref{sec:DIAG}, we have discussed the case of great
diagonal magnetic moments. This situation takes place when a transition
magnetic moment is suppressed by the GIM mechanism. If one considers
the $\nu_{e}\to\nu_{\tau}$ oscillations channel, i.e. relatively
small vacuum mixing angle, we can find the analytical transition probability
for spin-flavor oscillations of neutrinos with great diagonal magnetic
moments in matter and an electromagnetic wave; cf. Eqs.~(\ref{eq:PLRbigmua})
and~(\ref{eq:Aa}). However, the situation of great diagonal magnetic
moments is not very interesting from the point of view of phenomenology
since the GIM mechanism is valid if $\mu_{a}\sim m_{a}$~\cite{FukYan03}.
It makes $\mu_{a}$ to be very small for reasonable neutrino masses~\cite{Ase11}.
Therefore $A_{a}\ll1$ in Eq.~(\ref{eq:Aa}) and, hence, $P_{\beta\mathrm{L}\to\alpha\mathrm{R}}\ll1$
in Eq.~(\ref{eq:PLRbigmua}).

We have also considered the case of the great transition magnetic
moment in Sec.~\ref{sec:TRANS}. In this situation, we have derived
the effective Schr\"odinger Eq.~(\ref{eq:tildeH}) and have found its
exact solution for the $\nu_{e}\to\nu_{\tau}$ oscillations channel
neglecting $V$ in Eq.~(\ref{eq:tildeH}). Comparing Eqs.~(\ref{eq:PLRapprox})
and~(\ref{eq:A12}), as well as Eqs.~(\ref{eq:PLRbigmua}) and~(\ref{eq:Aa}),
with the analogous transition probability derived in Ref.~\cite{EgoLobStu00},
one can see that the results of Ref.~\cite{EgoLobStu00} are not
applicable for the description of neutrino spin-flavor oscillations
in the considered external fields. The reason for the discrepancy of our results and those in Ref.~\cite{EgoLobStu00} has been analyzed in Sec.~\ref{sec:TRANS}. Then, we have examined the numerical
solution of Eq.~(\ref{eq:tildeH}) and revealed that the obtained
exact solution qualitatively describes $\nu_{e\mathrm{L}}\to\nu_{\tau\mathrm{R}}$
oscillations.

Finally, basing on Eqs.~(\ref{eq:tildeH}) and~(\ref{eq:PLRnumsol}),
we have numerically studied $\nu_{e\mathrm{L}}\to\nu_{\mu\mathrm{R}}$
oscillations in matter with different densities. The transition probabilities
have been plotted in Fig.~\ref{fig:eLmuR}. One can see in Fig.~\ref{fig:eLmuR}
that, if one accounts for the high matter density in neutrino spin-flavor
oscillation in a plane electromagnetic wave, it diminishes the averaged
transition probability. Thus one does not expect the appearance of
a resonance in spin-flavor oscillations in the considered external
fields, predicted in Ref.~\cite{EgoLobStu00}.

At the end of this section, we mention that described neutrino spin-flavor
oscillations in background matter and a plane electromagnetic wave
can take place in the vicinity of a highly magnetized compact astrophysical
object, emitting intense electromagnetic radiation, being surrounded by
dense matter, and being a source of neutrinos. It can be, e.g., a
pulsar with a dense accretion disk. The estimates of the parameters
of the neutrino system and the external fields, corresponding to the
implementation of these spin-flavor oscillations in astrophysical
media, are given in Ref.~\cite{Dvo18} and Sec.~\ref{sec:TRANS}.

\section*{Acknowledgments}

This work was partially supported by RFBR (Grant No. 18-02-00149a).
I am also thankful to V.~G.~Bagrov for useful comments.

\end{document}